
%
\input harvmac

\def\CTPa{\it Center for Theoretical Physics, Department of Physics,
      Texas A\&M University}
\def\CTPb{\it College Station, TX 77843-4242, USA}
\def\HARCa{\it Astroparticle Physics Group,
Houston Advanced Research Center (HARC)}
\def\HARCb{\it The Woodlands, TX 77381, USA}

\def\CERN{\it CERN Theory Division, 1211 Geneva 23, Switzerland}
\def\ie{\hbox{\it i.e.}}     
\def\eg{\hbox{\it e.g.}}

\def\coeff#1#2{{\textstyle{#1\over #2}}}

\catcode`\@=11 

\def\lsim{\mathrel{\mathpalette\@versim<}}
\def\gsim{\mathrel{\mathpalette\@versim>}}
\def\@versim#1#2{\vcenter{\offinterlineskip
    \ialign{$\m@th#1\hfil##\hfil$\crcr#2\crcr\sim\crcr } }}
\def\boxit#1{\vbox{\hrule\hbox{\vrule\kern3pt
      \vbox{\kern3pt#1\kern3pt}\kern3pt\vrule}\hrule}}

\def\etal{{\it et. al.}}
\def\r#1{$\bf#1$}
\def\rb#1{$\bf\overline{#1}$}

\def\t1{{\tilde 1}}
\def\ov{\overline}

\def\JL{J. L. Lopez}
\def\DVN{D. V. Nanopoulos}

\def\GeV{\,{\rm GeV}}
\def\TeV{\,{\rm TeV}}

\def\wt{\widetilde}

\def\NPB#1#2#3{Nucl. Phys. B {\bf#1} (19#2) #3}
\def\PLB#1#2#3{Phys. Lett. B {\bf#1} (19#2) #3}

\def\PRD#1#2#3{Phys. Rev. D {\bf#1} (19#2) #3}
\def\PRL#1#2#3{Phys. Rev. Lett. {\bf#1} (19#2) #3}
\def\PRT#1#2#3{Phys. Rep. {\bf#1} (19#2) #3}

\def\TAMU#1{Texas A \& M University preprint CTP-TAMU-#1}

\nref\GG{H. Georgi and S. L. Glashow, \PRL{28}{72}{1494}.}
\nref\GWS{S. L. Glashow, \NPB{22}{61}{579}; S. Weinberg, \PRL{19}{67}{1264};
A. Salam, in Proceedings of the 8th Nobel Symposium, Stockholm 1968, ed.
N. Svartholm (Almqvist and Wiksells, Stockholm 1968) p. 367.}
\nref\ELN{J. Ellis, J. Lopez, and \DVN, \PLB{245}{90}{375};
A. Font, L. Ib\'a\~nez, and F. Quevedo, \NPB{345}{90}{389}.}
\nref\books{See \eg, {\it String theory in four dimensions}, ed.
by M. Dine, (North-Holland, Amsterdam, 1988);
{\it Superstring construction}, ed. by A.N. Schellekens
(North-Holland, Amsterdam, 1989).}
\nref\Barr{S. Barr, \PLB{112}{82}{219}, \PRD{40}{89}{2457}; J. Derendinger,
J. Kim, and \DVN, \PLB{139}{84}{170}.}
\nref\AEHN{I. Antoniadis, J. Ellis, J. Hagelin, and \DVN, \PLB{194}{87}{231}.}
\nref\JHW{I. Antoniadis, J. Ellis, J. Hagelin, and \DVN, \PLB{231}{89}{65};
\JL\ and \DVN, \PLB{268}{91}{359}; For a recent review see \JL\ and \DVN, in
Proceedings of the 15th Johns Hopkins Workshop on Current Problems in Particle
Theory, August 1991, p. 277; ed. by G. Domokos and S. Kovesi-Domokos.}
\nref\CAN{A. Chamseddine, R. Arnowitt, and P. Nath, \PRL{49}{82}{970}.}
\nref\DG{S. Dimopoulos and H. Georgi, \NPB{193}{81}{150}.}
\nref\IR{L. Ib\'a\~nez and G. Ross, \PLB{110}{82}{215}; \DVN\ and K. Tamvakis,
\PLB{110}{82}{449}.}
\nref\Lahanas{H. Nilles, M. Srednicki, and D. Wyler, \PLB{124}{83}{337};
A. Lahanas, \PLB{124}{83}{341}.}
\nref\MPM{A. Masiero, \DVN, K. Tamvakis, and T. Yanagida, \PLB{115}{82}{380};
B. Grinstein, \NPB{206}{82}{387}.}
\nref\Cosmo{C. Kounnas, \DVN, M. Quiros, and M. Srednicki, \PLB{127}{83}{82};
T. H\"ubsch, S. Meljanac, S. Pallua, and G. Ross, \PLB{161}{85}{122}.}
\nref\LN{For a review see A. Lahanas and \DVN, \PRT{145}{87}{1}.}
\nref\Witten{J. P. Derendinger, L. Ib\'a\~nez, and H. Nilles,
\PLB{155}{85}{65};
M. Dine, R. Rohm, N. Seiberg, and E. Witten, \PLB{156}{85}{55}.}
\nref\aspects{S. Kelley, \JL, \DVN, H. Pois, and K. Yuan, \TAMU{16/92} and
CERN-TH.6498/92.}
\nref\dfive{S. Weinberg, \PRD{26}{82}{287}; N. Sakai and T. Yanagida,
\NPB{197}{82}{533}.}
\nref\CEN{B. Campbell, J. Ellis, and \DVN, \PLB{141}{84}{229}.}
\nref\Japs{M. Matsumoto, J. Arafune, H. Tanaka, and K. Shiraishi,
University of Tokyo preprint ICRR-267-92-5 (April 1992).}
\nref\AN{R. Arnowitt and P. Nath, \TAMU{24/92}.}
\nref\LNY{\JL, \DVN, and K. Yuan, \NPB{370}{92}{445}.}
\nref\DM{S. Kelley, \JL, \DVN, H. Pois, and K. Yuan, in preparation.}
\nref\LNP{\JL, \DVN, and H. Pois, in preparation.}
\nref\Griest{K. Griest and D. Seckel, \PRD{43}{91}{3191};
P. Gondolo and G. Gelmini, \NPB{360}{91}{145}.}
\nref\KT{See \eg, E. Kolb and M. Turner, {\it The Early Universe}
(Addison-Wesley, 1990).}
\nref\inflation{For recent reviews see \eg, K. Olive, \PRT{190}{90}{307};
D. Goldwirth and T. Piran, \PRT{214}{92}{223}.}
\nref\Smoot{G. F. Smoot, \etal, COBE preprint 1992.}
\nref\Wright{E. L. Wright, \etal, COBE preprint 1992.}
\nref\susyinf{S. Hawking, \PLB{115}{82}{295}; A. Guth and S.-Y. Pi,
\PRL{49}{82}{1110}; J. Ellis, \DVN, K. Olive, and K. Tamvakis,
\PLB{120}{82}{331}.}
\nref\hybrid{R. Schaefer and Q. Shafi, Bartol preprint BA-92-28 (1992);
G. Efstathiou, J. R. Bond, and S. D. M. White, Oxford University preprint
OUAST/92/11.}
\nref\TRR{A. N. Taylor and M. Rowan-Robinson, Queen Mary College preprint,
June 1992.}
\nref\NO{\DVN\ and K. Olive, Nature {\bf327} (1987) 487.}
\nref\faspects{J. Ellis, J. Hagelin, S. Kelley, and \DVN, \NPB{311}{88/89}{1}.}
\nref\witt{E. Witten, \PLB{155}{85}{151}.}

\nfig\I{The fine-tuning coefficients $c_\mu$ and $c_t$ as a function of
$m_t$ for the five points in Table I (and $\tan\beta=1.73$; the sign of $\mu$
does not affect these coefficients) which satisfy the proton decay constraints
in the minimal $SU(5)$ supergravity model. Values of $c=\Delta$ indicate
a fine-tuning of model parameters of $\log(\Delta)$--orders of magnitude.}
\nfig\II{The lightest neutralino relic density as a function of $m_t$ for the
points in Table I (and $\tan\beta=1.73$ and both signs of $\mu$) which satisfy
the proton decay constraints in the minimal $SU(5)$ supergravity model.
(a) Values of $\Omega_\chi h^2>1$ (dashed line) are in conflict with current
cosmological observations; (b) small cosmologically allowed regions of
parameter space may still exist for special values of $m_t$. If $h=0.5$ then
$\Omega_\chi h^2>0.25$ (dotted line) is excluded.}

\Title{\vbox{\baselineskip12pt\hbox{CERN-TH.6554/92}\hbox{CTP--TAMU--49/92}
\hbox{ACT--14/92}}}
{\vbox{\centerline{Troubles with the Minimal SU(5) Supergravity Model}}}
\centerline{JORGE~L.~LOPEZ$^{(a)(b)}$, D.~V.~NANOPOULOS$^{(a)(b)(c)}$,
and A. ZICHICHI$^{(d)}$}
\smallskip
\centerline{$^{(a)}$\CTPa}
\centerline{\CTPb}
\centerline{$^{(b)}$\HARCa}
\centerline{\HARCb}
\centerline{$^{(c)}$\CERN}
\centerline{$^{(d)}${\it CERN, Geneva, Switzerland}}
\vskip .1in
\centerline{ABSTRACT}
We show that within the framework of the minimal $SU(5)$ supergravity model,
radiatively-induced electroweak symmetry breaking and presently available
experimental lower bounds on nucleon decay, impose severe constraints on the
available parameter space of the model which correspond to fine-tuning of the
model parameters of over two orders of magnitude. Furthermore, a
straightforward calculation of the cosmic relic density of neutralinos ($\chi$)
gives $\Omega_\chi h^2\gg1$ for most of the allowed parameter space in this
model, although small regions may still be cosmologically acceptable. We
finally discuss how the {\it no-scale flipped $SU(5)$ supergravity model}
avoids naturally the above troubles and thus constitutes a good candidate for
the low-energy effective supergravity model.
\Date{June, 1992}

\newsec{Introduction}
Mathematical simplicity or economy does not always imply physical simplicity
or entails on nature what choices to make. A typical example is electroweak
unification, where the simplest possible mathematical choice which encompasses
charged ($W^\pm$) and electromagnetic ($\gamma$) currents is $SU(2)$ \GG.
But we all know that nature's preferred choice is $SU(2)\times U(1)$ \GWS.
One of the best reasons we can offer is the availability of non-vector
representations in $SU(2)\times U(1)$. In the case of $SU(2)$ there is no
reason for the fermion masses to be $\lsim{\cal O}(M_W)$, whereas in
$SU(2)\times U(1)$ at least we understand naturally why this must be so.

The point we would like to make here is that one can either follow a big
fundamental principle to choose {\it the model} and/or follow indicative
clues from the available experimental data.  String theory may eventually
provide us with a unique vacuum (\ie, model) and then all discussions would
be hushed. For the time being though, we have to struggle between the two
above mentioned ways to select the potentially right model.

Daring minds have used the absence of adjoint Higgs representations \ELN\
in level-one Kac-Moody superstring constructions \books\ as a super-clue to
select flipped $SU(5)\times U(1)$ \refs{\Barr,\AEHN} as the ``chosen" model
\JHW. For those who feel queasy by the
presence of the extra $U(1)$ factor, we remind them of the opening paragraph
(\ie, $SU(2)$: out; $SU(2)\times U(1)$: in) and of the fact that in string
theories $U(1)$ factors proliferate.

Here we will follow well established theoretical prejudices (radiative
electroweak breaking) and available lower bounds on the nucleon decay
lifetime to corner the minimal $SU(5)$ supergravity model, and then use
calculations of the cosmic neutralino relic density (which turns out to be
almost always very large $\Omega_\chi\gg1$), to cast doubts about the
candidacy of this model as the effective low-energy supergravity model.

The minimal $SU(5)$ supergravity model \CAN\ can be generically described by
the
following observable sector superpotential
\eqn\I{W_o=\lambda_1(\coeff{1}{3}\Sigma^3+\coeff{1}{2}M\Sigma^2)
+\lambda_2(h\bar h\Sigma+3M'h\bar h)+\lambda_3h\bar h\phi +\lambda_t FFh
+\lambda_b F\bar f\bar h,}
where $\Sigma$ is the \r{24} of $SU(5)$ whose scalar components acquire
a vev
$\Sigma_{xy}=M[2\delta_{xy}-5(\delta_{5x}\delta_{5y}+\delta_{4x}\delta_{4y})]$
which breaks $SU(5)$ down to $SU(3)\times SU(2)\times U(1)$; $h,\bar h$ are
\r{5},\rb{5} Higgs superfields; $\phi$ is a singlet which induces
$SU(2)_L\times U(1)_Y\to U(1)_{em}$ (tree-level) breaking although it can be
omitted if the electroweak symmetry is broken radiatively; and $F,\bar f$ are
the usual \r{10},\rb{5} matter fields. The most pervasive difficulty
encountered
in this model is the needed doublet-triplet mass splitting of the Higgs
pentaplets. As is, in Eq. \I\ the choice $M'=M$ gives massless doublets and
${\cal O}(M)$ triplets \DG. Even though stable under radiative corrections,
this
solution is rather ad-hoc. A more natural solution is provided by the `sliding
singlet mechanism' \IR\ in which the $M'h\bar h$ term is dropped but the
singlet
$\phi$ gets a large vev which again keeps the doublets massless. In either case
the remaining physical degrees of freedom of the singlet are light, and after
supersymmetry breaking the Higgs doublets become massive as follows. The
$h\bar h\phi$ term gives rise to supersymmetry breaking trilinear scalar
couplings $A\phi H_2\bar H_2$ and $A\phi H_3\bar H_3$, and $h\bar h\Sigma\to
M_U H_3\bar H_3$ to a soft supersymmetry breaking scalar mass squared
$\sim M_U\wt m$ which splits the $H_3$ supermultiplet (which acquires
an average mass $\sim M_U$). These three couplings generate a one-loop $H_2$
self-energy tadpole diagram mediated by $\phi$ and having $H_3$ circulating in
the loop, which gives a $\sim(M_U\wt m)^{1/2}$ mass to the scalar components of
$H_2,\bar H_2$ \Lahanas. Thus the light singlet and the large Higgs triplet
mixing destroy the doublet-triplet splitting.

The best known solution to this problem utilizes the `missing partner
mechanism' \MPM\ wherein a \r{75} of $SU(5)$ breaks the gauge symmetry and new
\r{50},\rb{50} representations are introduced: the three $h\bar h$ terms
in Eq. \I\ are replaced by ${\bf50}\cdot{\bf75}\cdot h$ and
$\ov{\bf50}\cdot{\bf75}\cdot \bar h$. The doublets now remain massless
automatically (since there are no doublets in the \r{50},\rb{50}),
whereas the triplets acquire $\sim M_U$ masses. Unfortunately,
the $SU(5)$ gauge symmetry breaking through a vev of the \r{75} causes
several cosmological difficulties \Cosmo.

Let us now turn to the hidden sector superpotential which gives rise to
soft supersymmetry breaking masses and couplings upon spontaneous breakdown
of supergravity. In the minimal supergravity model there is a problem with
the vacuum energy which is generically $\sim M^4_{Pl}$ after supergravity
breaking, unless one arranges the parameters in the hidden superpotential
to obtain zero vacuum energy. This is a fine-tuning which is not protected
by any symmetry of the theory. This problem is solved by going to no-scale
supergravity \LN\ where the vacuum energy is automatically zero even after
supersymmetry breaking. Another problem of the hidden sector in the minimal
model is that the scale of supersymmetry breaking must be put in by hand as one
of the parameters and there is no justification for the phenomenologically
acceptable choice. In contrast, no-scale supergravity owes its name to the
fact that in this class of theories it may be possible to determine the
magnitude of the supersymmetry breaking parameters dynamically. Another
solution
to this problem of the hidden sector of the minimal model is obtained in
dynamical models of supersymmetry breaking \Witten\ where this scale is related
to the scale where hidden sector gaugino condensates form.

\newsec{Proton decay and fine-tuning}
The parameter space of unified supergravity models with universal soft
supersymmetry breaking can be described by just five parameters: universal
scalar ($m_0$) and gaugino ($m_{1/2}$) masses, universal cubic scalar couplings
($A$), the ratio of the Higgs vacuum expectation values ($\tan\beta$), and
the top-quark mass ($m_t$). The remaining parameters in the model are
determined
from the experimental values of $M_Z,m_b,m_\tau,\alpha_3,\alpha_{em}$, the
renormalization group equations, and the constraint of radiative electroweak
symmetry breaking. (The sign of the Higgs mixing parameter $\mu$ remains
undetermined.) A detailed study of this parameter space using the one-loop
Higgs effective potential has been recently given in Ref. \aspects. The
five-dimensional parameter space is restricted by several consistency and
phenomenological constraints. This space is bounded in the $\tan\beta$ and
$m_t$ directions but there is no firm upper bound on the soft supersymmetry
breaking parameters $m_{1/2}$, $\xi_0\equiv m_0/m_{1/2}$, and
$\xi_A=A/m_{1/2}$.
A semi-quantitative upper bound on $m_{1/2}$ (as a function of $\xi_0$) can
be obtained by demanding a `not-too-much-fine-tuning' condition. In Ref.
\aspects\ two fine-tuning coefficients $c_\mu$ and $c_t$ have been defined
which have the property that if $c_{\mu,t}<\Delta$ then the observed value of
$M_Z$ is derived within the model with cancellations among the relevant
parameters of less than $\log(\Delta)$--orders of magnitude. The physical
meaning of these coefficients can be readily grasped by studying the following
tree-level Higgs potential minimization constraint
\eqn\II{\coeff{1}{2}M^2_Z={m^2_{H_1}-m^2_{H_2}\tan^2\beta\over \tan^2\beta-1}
-\mu^2,}
where $m_{H_{1,2}}$ are the Higgs soft supersymmetry breaking masses. Since
the whole theory (the Higgs potential and the RGEs) scales with $m_{1/2}$,
as $m_{1/2}$ grows the first term on the right-hand-side will tend to grow
forcing $\mu$ to larger values to keep $M_Z$ at its observed value. This
results in large values of
$c_\mu\equiv|(\mu^2/M^2_Z)\partial M^2_Z/\partial\mu^2|=2\mu^2/M^2_Z$. It
could also happen
that $\mu$ is kept artificially small by some cancellations within the first
term on the right-hand-side of Eq. \II\ due to fine-tuned choices of $m_t$.
This results in large values of
$c_t\equiv|(m^2_t/M^2_Z)\partial M^2_Z/\partial m_t^2|$. The main result is
that generically $c_{\mu,t}\propto m^2_{1/2}(a+b\xi^2_0)$, where $a$ and $b$
are some $m_t$- and $\tan\beta$-dependent functions. The latter generally
increases with $m_t$. In Ref. \aspects\ it was found that for
$m_{1/2}\gsim400\GeV$, $c_{\mu,t}\gsim100$ for all values of $\xi_0$. The
choice $\Delta=100$ thus corresponds to the usual `naturalness' bound of
$m_{\tilde g}=2.77 m_{1/2}\lsim1\TeV$. Smaller values of $m_{1/2}$ allow some
range of values of $\xi_0$ such that $c_{\mu,t}<\Delta$.

The above analysis has been performed in Ref. \aspects\ for the supersymmetric
standard model (SSM), \ie, a generic unified supergravity model which reduces
to the minimal
supersymmetric standard model (MSSM) at low energies. In the specific case of
the minimal $SU(5)$ supergravity model one must ensure that the dimension-five
proton-decay-mediating operators which result by integrating out the
$H_3,\bar H_3$ Higgs triplets \dfive, give values of the $p\to\bar\nu K^+$
decay rate which are compatible with the current experimental bound. This
requirement imposes severe constraints on the parameter space of the model
\refs{\CEN,\Japs,\AN}. Without getting into the calculational details, one
can simply state that this requirement implies an upper bound on a quantity
$P=P(m_{1/2},\xi_0,\xi_A,m_t,\tan\beta)<(103\pm15)M_{H_3}/M_U$ \AN. The
function
$P$ is reduced by small values of $m_{1/2}$ and large values of $\xi_0$, and
it grows with $\tan\beta$, $P\propto(1+\tan^2\beta)/\tan\beta$. If one assumes
that $M_{H_3}<3M_U$ so that the Yukawa coupling which gives rise to the $H_3$
mass is reasonably small \AN, then an absolute upper bound on $P$ results. In
Ref. \AN\ the resulting allowed parameter space is given for values of
$m_t=125\GeV$ and $\tan\beta=1.73$ and a calculable value of $\xi_A$. The
values of $m_{1/2}$ and $m_0$ are varied over a range which presumably respects
the `naturalness' constraints. We have selected five
$(m_{1/2},\xi^{min}_0,\xi_A)$ sets (given in Table I)
which are on the boundary of the allowed region (\ie, on the horizontal line
in Fig. 2 of Ref. \AN). Values of $\xi_0$ smaller than these (for fixed
$m_{1/2}$) give values of $P$ inconsistent with the experimental upper bound.
For the purposes of this paper the choice $\xi_0=\xi^{min}_0$ will suffice.
For $\tan\beta=1.73$ we have computed $c_\mu$ and $c_t$ for varying values of
$m_t$ for these five points,\foot{The values of $m_t$ vary over the shown
finite range which is completely determined by the choices of the other four
parameters \aspects. Also, all calculations in this paper are performed using
the one-loop Higgs effective potential.} as shown in Fig. 1. (These
coefficients do not depend on
the sign of $\mu$.) It is clear that $c_t$ exceeds $\Delta=100$ for all values
of $m_t$ and for all the selected points. Since $c_{\mu,t}$ grow with
$\xi^2_0$, values of $\xi_0>\xi^{min}_0$ (\ie, inside the allowed region) give
even larger values of $c_{\mu,t}$. Furthermore, if we take $M_{H_3}=M_U$ as a
reasonable central value for $M_{H_3}$, then the upper bound on $P$ gets
reduced by a factor of 3 and the values of $\xi^{min}_0$ grow significantly.
For example, for $m_{1/2}=74\,(122)\GeV$ (\ie, curve a (b) in Fig. 1),
$\xi^{min}_0$ goes from $8.1\,(6.5)$ to $16.2\,(11.5)$. The values of
$c_{\mu,t}$ grow accordingly; for $m_t=125\GeV$, $c_t$ goes from $\approx100\,
(200)$ to $\approx400\,(600)$.

\topinsert
\noindent {\bf Table I}: Selected set of points which are on the boundary of
the allowed region which satisfies proton decay constraints for $m_t=125\GeV$,
$\tan\beta=1.73$, and $\mu>0$, and the corresponding gluino and average squark
masses, fine-tuning coefficients, and neutralino relic density.
Points inside the allowed region have $\xi_0>\xi^{min}_0$ and larger values of
$m_{\tilde q}$ and $\Omega_\chi h^2$. The `symbol' corresponds to the curves
shown in Figures 1 and 2. All masses in GeV.
\medskip
\input tables
\thicksize=1.0pt
\centerjust
\begintable
symbol|$m_{1/2}$|$\xi^{min}_0$|$\xi_A$|$m_{\tilde g}$|$m_{\tilde q}$
|$c_\mu$|$c_t$|$\Omega_\chi h^2$\cr
a|$74$|$8.1$|$-5.4$|$205$|$625$|$33$|$121$|$15.9$\nr
b|$122$|$6.5$|$-3.2$|$340$|$845$|$53$|$218$|$27.9$\nr
c|$187$|$5.4$|$-1.6$|$520$|$1110$|$78$|$364$|$5.13$\nr
d|$267$|$4.5$|$-.41$|$740$|$1370$|$104$|$530$|$3.74$\nr
e|$364$|$3.8$|$+.55$|$1010$|$1645$|$133$|$726$|$3.20$\endtable
\endinsert

\newsec{The neutralino relic density}
The parameter space of the minimal $SU(5)$ supergravity model is highly
constrained by the experimental proton decay bounds as discussed above.
We now show that for the allowed region the cosmic relic density of the
lightest supersymmetric particle (LSP), \ie, the lightest neutralino, is
generally very large, \ie, $\Omega_\chi h^2\gg1$, where $0.5\le h\le1$ is the
Hubble parameter.

We have calculated $\Omega_\chi h^2$ in this model for the five points given in
Table I following the methods described in Ref. \LNY. This is a numerically
intensive calculation which differs from the usual analyses in that in the
computation of the LSP annihilation cross section the masses and couplings
of all particles involved can be determined for any choice of the five model
parameters. That is, no ad-hoc assumptions are made about the masses of the
exchanged and final-state particles. The present calculation includes the
one-loop corrected Higgs boson masses as well.\foot{A detailed study of the
cosmic relic density of neutralinos in the SSM is given in Ref. \DM.}

The results depend on the sign of $\mu$ and are given in Fig. 2 as a function
of $m_t$ for the five points in Table I and $\tan\beta=1.73$. These are
actually lower bounds since $\xi_0>\xi^{min}_0$ increases $\Omega_\chi h^2$.
Recall also that $M_{H_3}<3 M_U$ leads to larger values of $\xi^{min}_0$.
In all honesty, only the values of $\Omega_\chi h^2$ for $\mu>0$ and
$m_t=125\GeV$ (\ie, those given in Table I) can be used to further constrain
the allowed parameter space. This is so because for other values of $m_t$
and/or the sign of $\mu$, the value of the function $P$ (see Sec. 2) will be
different from the one which follows for the points in Table I, \ie, $P=309$
\AN. A detailed analysis of these cosmological bounds on the minimal $SU(5)$
supergravity model will be given elsewhere \LNP.

It is clear that generally $\Omega_\chi h^2\gg1$ (see Fig. 2a) unless $m_t$
takes `special' values as shown in detail in Fig. 2b. The explanation for this
phenomenon is simple: for values of $m_t$ close to the low end of their
allowed range, $\mu$ is relatively small (see Fig. 1, keeping in mind that in
tree-level approximation $c_\mu=2\mu^2/M^2_Z$), and the LSP composition
is `mixed', allowing for `normal' levels of annihilation, and therefore
small values of $\Omega_\chi h^2$ \LNY. Note though that since the squarks,
sleptons, and the three heavier Higgs bosons are rather heavy due to the proton
decay constraints \AN, only the $Z$ and lightest Higgs $h$ remain as efficient
annihilation mediators. When $m_t$ (and therefore $\mu$) grows, the LSP becomes
increasingly more a nearly pure bino state (see \eg, Fig. 1 in Ref. \LNY)
and its couplings to $Z$ and $h$ tend to zero \LNY, resulting in a large relic
density.

Some of the curves (curve `a' for $\mu>0$ and curves `b' and `c' for $\mu<0$)
exhibit a non-monotonic dependence of $\Omega_\chi h^2$ with $m_t$. This is due
to poles and thresholds of the LSP annihilation cross section\foot{Our
calculational scheme breaks down for points near these special regions;
detailed calculations show that the correct result varies more smoothly than
our figures indicate \Griest.} for special values of $m_\chi$. For example, in
curve `b' for $\mu<0$, at $m_t\approx111\GeV$,
$m_\chi\approx{1\over2}m_h=33\GeV$, and at $m_t\approx122\GeV$,
$m_\chi\approx{1\over2}M_Z$. The spike on curve `c' for $\mu<0$ and
$m_t\approx112\GeV$ is due to the newly openned $\chi\chi\to hh$ channel for
$m_\chi=m_h\approx60\GeV$.

Let us now address the question of what values of $\Omega_\chi h^2$ are
excluded on cosmological grounds. There is no observational evidence for
astronomical systems with $\Omega\gg1$ \KT. In fact, most studies indicate that
$\Omega<1$. On phenomenological grounds, cold dark matter seeded structure
formation models require $\Omega\le1$ \KT. On theoretical grounds $\Omega=1$
is the only `time-stable' value, in that smaller values must be fine-tuned to
be very close to unity otherwise the Universe would have re-collapsed on a
Planck time scale. Inflationary models of course predict this precise value of
$\Omega$ \inflation.

Recent data from the COBE DMR instrument \Smoot\ showing a non-vanishing
quadrupole moment of the cosmic microwave background anisotropy appear to
confirm the basic predictions of inflationary models \Wright.\foot{In fact,
the needed small density perturbations $\delta\rho/\rho={\cal O}(10^{-5})$
are only compatible with {\it supersymmetric} inflationary models \susyinf.}
Furthermore, the `best fit' to the data seems to be given by a mixture of cold
and hot dark matter \refs{\hybrid,\TRR}, as originally proposed in Ref. \NO,
once more disfavoring values of $\Omega>1$. Moreover, $h=0.5$ appears to be
strongly favored over $h=1$ \TRR, in which case values of
$\Omega_\chi h^2>0.25$ are disfavored on cosmological grounds.

Table I then shows that the particular set of representative points chosen
which satisfy the proton decay constraints in this model are in gross conflict
with cosmological observations. It may be possible to find small regions of
parameter space where the value of $m_t$ is tuned (as Fig. 2b shows) to be
within narrow intervals such that $\Omega_\chi h^2<1$.  Note that there are
several obstacles hampering the identification of these possible cosmologically
allowed regions: (i) values of $\xi_0>\xi^{min}_0$ increase $\Omega_\chi h^2$,
(ii) values of $M_{H_3}<M_U$ increase $\xi^{min}_0$ and therefore
$\Omega_\chi h^2$, (iii) $h=0.5$ decreases the allowed region considerably
since in this case $\Omega_\chi h^2>0.25$ is excluded.

\newsec{Discussion}
We have shown that proton decay constraints force the minimal $SU(5)$
supergravity model into a region of parameter space where the $Z$-mass is
obtained within this model subject to cancellations among the model parameters
of at least two orders of magnitude. Furthermore, within this allowed region
the relic density of neutralinos is generally in gross conflict with current
cosmological observations, although small regions of parameter space may still
exist which are cosmologically acceptable. In our opinion, these results cast
doubts about the candidacy of the minimal $SU(5)$ supergravity model
as the correct low-energy effective supergravity model.

We now present an alternative supergravity model based on no-scale supergravity
with the gauge group flipped $SU(5)$ which does not suffer from any of the
troubles discussed in this paper. The doublet-triplet splitting of the Higgs
pentaplets is achieved through the $SU(5)$ invariant couplings $HHh$ and
$\bar H\bar H \bar h$ \AEHN, where $H,\bar H$ are \r{10},\rb{10} Higgs
representations which effect the $SU(5)\times U(1)\to SU(3)\times SU(2)\times
U(1)$ symmetry breaking when their neutral components acquire non-zero vevs.
This is possible because the distribution of the quarks and leptons in the
usual \rb{5} and \r{10} representations is `flipped' (\ie, $u\leftrightarrow
d$,
$e\leftrightarrow\nu$) relative to their usual assignments, and therefore
$H\supset\nu^c_H$, $\bar H\supset\nu^c_{\bar H}$.\foot{The fact that symmetry
breaking does not require adjoint Higgs representations is crucial to the
derivation of flipped $SU(5)$ models from superstring theory \ELN.} The
above couplings then give $HHh\to M_U d^c_H H_3$ and $\bar H\bar H\bar h\to
M_U d^c_{\bar H} \bar H_3$, making the triplets heavy and leaving the doublets
massless. Note that this pattern of symmetry breaking is unique \AEHN, thus
avoiding the cosmological multiple-vacua problem of regular $SU(5)$.

As far as the dimension-five proton decay operators are concerned, note that
the Higgs triplet mixing term $h\bar h\phi\to M_W H_3\bar H_3$ ({\it c.f.}
$h\bar h\Sigma\to M_U H_3\bar H_3$) is small, whereas the triplet masses
themselves are large. This results in a $M_W/M_U$ suppression (in the
amplitude)
of these operators relative to the regular $SU(5)$ case \faspects, thus making
them completely negligible. This implies that values of the supersymmetry
breaking parameters do not need to be as large and therefore the fine-tuning
coefficients can be naturally small. Furthermore, $\Omega_\chi h^2$ can be
within current cosmological bounds for a wide range of model parameters,
perhaps even providing interesting amounts of astrophysical dark matter \DM.
The small Higgs triplet mixing term also prevents the light $H_2,\bar H_2$
doublets from acquiring large masses through the one-loop tadpole diagram
discussed in the Introduction. Indeed, in the case of flipped $SU(5)$ the
induced $H_2$ scalar mass is $\sim(M_W\wt m)^{1/2}\sim M_W$.

Let us finally remark that no-scale supergravity ameliorates considerably
the cosmological constant problem, \ie, $\Lambda\sim M^4_W$, whereas
$\Lambda\sim M^4_{Pl}$ in minimal supergravity. Also, superstring models
realize the no-scale ansatz automatically \witt, and interesting flipped
$SU(5)$ models have been constructed within this framework \JHW. We thus
propose the no-scale flipped $SU(5)$ supergravity model as a very good
candidate for the low-energy effective supergravity model.

\bigskip
\bigskip
\bigskip
\noindent{\it Acknowledgments}: We would like to thank H. Pois and K. Yuan
for useful discussions. This work has been supported in part by DOE
grant DE-FG05-91-ER-40633. The work of J.L. has been supported in part by an
ICSC-World Laboratory Scholarship. The work of D.V.N. has been supported in
part by a grant from Conoco Inc. We would like to thank the
HARC Supercomputer Center for the use of their NEC SX-3 supercomputer.

\listrefs
\listfigs
\bye